\documentclass{article} % For LaTeX2e
\usepackage{iclr2026_conference,times}

\usepackage{hyperref}
\usepackage{url}
\usepackage{booktabs}
\usepackage{graphicx}
\usepackage{amsmath}
\iclrfinalcopy
\title{Compute Allocation for Reasoning-Intensive Retrieval Agents}

\author{Sreeja Apparaju \\
\texttt{sapparaju23@gmail.com} \\
\And
Nilesh Gupta \\
\texttt{nileshgupta2797@utexas.edu}
}

\begin{document}

\maketitle

\begin{abstract}
% Reasoning-intensive information retrieval (RIIR) is a core capability for agent memory, yet it is unclear how to optimally allocate finite inference compute across the retrieval stack to support complex reasoning. We adapt recent findings from RIIR to agent memory architectures, conducting systematic ablations of compute–quality trade-offs using the BRIGHT benchmark and the Gemini 2.5 model family. We decompose the memory retrieval pipeline into query expansion, initial retrieval, and LLM-based re-ranking, and systematically vary three knobs: model strength, inference-time “thinking” depth, and re-ranking depth. We propose a resource-efficient pipeline that concentrates compute on deep re-ranking with stronger models while utilizing lightweight models for query generation, achieving 93\% of maximum quality at 3.6x lower cost. Our findings establish concrete design principles for production retrieval systems and demonstrate that strategic compute allocation and not uniform scaling is the key to cost-effective agent memory.
As agents operate over long horizons, their memory stores grow continuously, making retrieval critical to accessing relevant information. Many agent queries require reasoning-intensive retrieval, where the connection between query and relevant documents is implicit and requires inference to bridge. LLM-augmented pipelines address this through query expansion and candidate re-ranking, but introduce significant inference costs.
We study computation allocation in reasoning-intensive retrieval pipelines using the BRIGHT benchmark and Gemini 2.5 model family. We vary model capacity, inference-time thinking, and re-ranking depth across query expansion and re-ranking stages. We find that re-ranking benefits substantially from stronger models (+7.5 NDCG@10) and deeper candidate pools (+21\% from $k$=10 to 100), while query expansion shows diminishing returns beyond lightweight models (+1.1 NDCG@10 from weak to strong). Inference-time thinking provides minimal improvement at either stage. These results suggest that compute should be concentrated on re-ranking rather than distributed uniformly across pipeline stages.
\end{abstract}

\section{Introduction}

Long-running agents accumulate memories containing past interactions, decisions, and task outcomes \citep{park2023generative, packer2023memgpt}. Retrieving relevant information from this memory is essential in grounding new decisions in prior experience. However, many retrieval queries in agent settings are \emph{reasoning-intensive}: the relevant information does not share lexical or semantic overlap with the query and can only be identified through inference.

Consider an agent asked \textit{``What task can I assign to the summer intern?''}. Relevant memories might include list of projects like \textit{``Project Alpha (requires security clearance)''} and \textit{``Project Gamma (exploratory research with a flexible timeline).''} Neither document mentions ``intern'' or ``task assignment'' yet both are directly relevant; though the first rules out, while the second suggests a good fit. Identifying this relevance requires reasoning about implicit constraints: interns lack security clearance and benefit from flexible, learning-oriented work. Embedding-based retrieval, which relies on surface similarity, cannot bridge this gap.
\begin{figure}[h]
\centering
\includegraphics[width=0.9\linewidth]{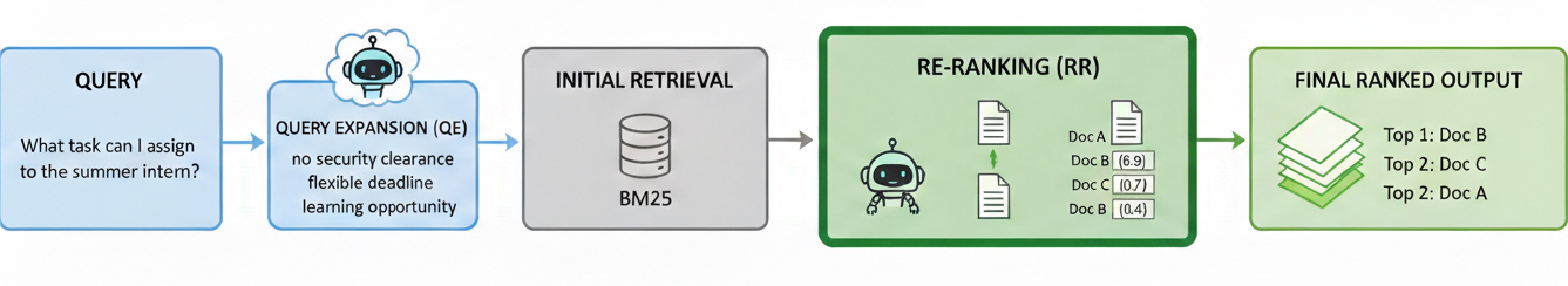}
\caption{Example of simplified LLM-augmented retrieval pipeline for reasoning-intensive queries}
\label{fig:intro_example}
\end{figure}

LLM-augmented retrieval pipelines address this limitation. A query expansion stage uses an LLM to surface implicit constraints \citep{gao2023hyde, wang2023query2doc}, improving recall. A re-ranking stage then evaluates retrieved candidates against the original query, recognizing relevance that keyword matching misses \citep{sun2023rankgpt}. Prior work on the BRIGHT benchmark \citep{su2025bright} demonstrates that such pipelines substantially outperform standard retrieval on reasoning-intensive queries. However, LLM inference at each stage introduces significant costs, raising a practical question: under a fixed compute budget, how should resources be allocated across pipeline stages?

We present a systematic study of compute allocation in LLM-augmented retrieval. Using BRIGHT as a testbed for reasoning-intensive retrieval, we vary three axes across Gemini 2.5 models: model capacity, inference-time thinking, and re-ranking depth. Our findings:
\begin{itemize}
    \item Query expansion shows diminishing returns: lightweight models capture most gains, with only 1.1 NDCG@10 improvement when scaling from weak to strong models.
    \item Re-ranking benefits substantially from stronger models (+7.5 NDCG@10) and deeper candidate pools (+21\% from $k$=10 to 100).
    \item Inference-time ``thinking'' provides minimal improvement at either stage.
\end{itemize}
These results indicate that compute should be concentrated on re-ranking rather than distributed uniformly across pipeline stages.

\section{Methodology}

\subsection{Pipeline Architecture}

Whether implemented via vector databases, lexical indices, or hybrid systems, we can model the agent's memory retrieval process for a reasoning intensive task as a three stage pipeline: 

\begin{itemize}
    \item \textbf{Query Expansion (QE):} An LLM expands the raw query $q$ to uncover implicit constraints and domain-specific terms ($q_{exp}=LLM_{\theta}(q)$)
    \item \textbf{Initial Retrieval:} An algorithm/agent to retrieve a candidate list $\mathcal{L}_{init}$ from a large corpus. We choose BM25 algorithm as it can adapt to different queries given that LLM-generated queries out-of-distribution for trained models. 
    \item \textbf{List-wise Re-ranking (RR):} An LLM examines the resulting top-$k$ candidates from initial retrieval to reason over their relevance and generate a finalized ranked list ($\pi_{top10}=LLM_{\phi}(q,\{d_{1},...,d_{k}\})$)
\end{itemize}

We use the Gemini 2.5 model family to represent a spectrum of cost and capability:
\begin{itemize}
\item \textbf{Model Strength:} \texttt{flash-lite} (highly efficient, no thinking), \texttt{flash-no-think} (mid-sized, thinking disabled), \texttt{flash-think} (mid-sized with dynamic thinking), and \texttt{pro} (large model with thinking) across QE and RR stages
\item \textbf{Thinking Depth:} Standard vs. extended reasoning modes (No-Think vs. Think)
\item \textbf{Re-ranking Depth:} $k \in \{10, 20, 50, 100\}$ documents
\end{itemize}

\textbf{Task \& Benchmark}: BRIGHT dataset that contains queries across 12 domains and requires multi-hop reasoning and implicit constraint satisfaction.

\textbf{Metrics}: NDCG@10 (ranking quality), Recall@10 (retrieval ceiling), cost per query (\$), and latency (seconds).

\section{Results}

\subsection{Scaling in Query Expansion (QE)}

Query expansion serves as an entry point for injecting parametric knowledge into the retrieval stack. For each query in the BRIGHT subsets, we generate an expanded query across four Gemini 2.5 models. Then, we perform BM25 retrieval using these "smarter" expanded queries, measuring NDCG@10 and Recall@10 to determine if high-capacity models provide a significantly better foundation for downstream re-ranking.
 Table~\ref{tab:qe_results} reveals a pattern of rapidly diminishing returns.

\begin{table}[h]
\centering
\caption{BM25 retrieval with different QE Model Variants}
\label{tab:qe_results}
\small
\begin{tabular}{@{}lccc@{}}
\toprule
\textbf{Model Variant} & \textbf{NDCG@10} & \textbf{Recall@10} & \textbf{Cost (\$/query)} \\ \midrule
No Enhancement (BM25 only)    & 14.52            & 33.76               & 0.000              \\
Flash-Lite        & 28.87            & 57.19               & 0.0018             \\
Flash (No-Think)  & 29.63            & 58.56               & 0.0093             \\
Flash (Think)     & 30.23            & 57.73               & 0.0141             \\
Pro               & 30.01            & 58.01               & 0.0489             \\ \bottomrule
\end{tabular}
\end{table}

The initial transition from raw queries to LLM-augmented expansion (Flash-Lite) produces dramatic gains: +14.35 NDCG@10 and +23.43 Recall@10. However, further investment yields negligible returns. Moving from Flash-Lite to a 27$\times$ more expensive Pro model improves Recall@10 by merely 0.82 points, a marginal return of 0.03 Recall points per dollar.

\textbf{Counterintuitive finding:}  Flash (Think) produces a lower Recall@10 than Flash (No-Think) (57.73 vs 58.56) despite a 52\% higher cost. This suggests that query expansion is fundamentally a vocabulary coverage task, rather than a reasoning challenge. Additional deliberation may cause the model to overthink straightforward lexical expansions and with increased inference-time reasoning actually \textit{degrades} performance. 

\subsection{Scaling in Re-ranking (RR)}
Re-ranking is where the system performs deliberative selection over retrieved candidates, and it is inherently more compute-intensive because it processes multiple documents per query. This segment introduces a multi-variable optimization problem within the pipeline, as performance is contingent upon the quality of the initial candidate set, the depth of the re-ranking pool (k), and the reasoning capacity of the re-ranker itself. 

\subsubsection{Ablation I: Varying Re-ranking Model Variant}

The “thinking” mode allows models to generate internal chain-of-thought reasoning before producing results. Unlike query expansion, where thinking showed limited value, re-ranking is a more complex task requiring careful comparison of multiple documents against nuanced query requirements. Here, we fix $k=100$ and QE to flash-think while varying the re-ranking model variant, isolating the impact of decision module’s strength quality at a fixed candidate coverage.

\begin{table}[h]
\centering
\caption{Impact of Re-ranker Model Strength ($k=$100; $QE=$ Flash No-Think)}
\label{tab:model_strength_rr}
\small
\begin{tabular}{@{}lcccccc@{}}
\toprule
\textbf{Configuration} & \textbf{NDCG@10} & \textbf{Recall@10} & \textbf{Cost (\$/query)} & \textbf{Latency (s/query)}\\ \midrule
QE (Flash) + RR (Flash-Lite)  & 36.54  & 38.76  & 0.021 & 20.25    \\
QE (Flash) + RR (Flash-no-think)       & 39.36  & 41.11  & 0.045  & 45.31   \\
QE (Flash) + RR (Flash-Think) & 39.92  & 42.03  & 0.054  & 63.28\\
QE (Flash) + RR (Pro)         & 41.19  &  43.57  &0.17 &79.25  \\ \bottomrule
\end{tabular}
\end{table}

Empirical analysis reveals that re-ranking is the primary driver of quality and exhibits far greater sensitivity to model capacity than query expansion. The pro model achieves the highest retrieval quality (41.19 NDCG@10), but flash-no-think emerges as the most cost-effective configuration, delivering 95.6\% of pro's performance at 3.8$\times$ lower cost. 

\textbf{Counterintuitive finding:} Enabling the “thinking” mode within the Flash model provides limited incremental benefit. Flash (Think) improves NDCG@10 only marginally over Flash (No-Think) (39.92 vs. 39.36) while incurring substantial overhead (20\% higher cost and 40\% higher latency).

\subsubsection{Ablation II: Varying Re-ranking Depth}
We fix the re-ranking model to Gemini-2.5-Flash (Think) and vary $k \in \{10, 20, 50, 100\}$ to isolate the effect of depth in ranking by examining $topk$ candidates. The central question is whether increasing the size of the candidate pool provides a reliable mechanism for surfacing relevant evidence that maybe buried deep in the initial retrieval results.

\begin{figure}[t]
\centering
\includegraphics[width=0.85\linewidth]{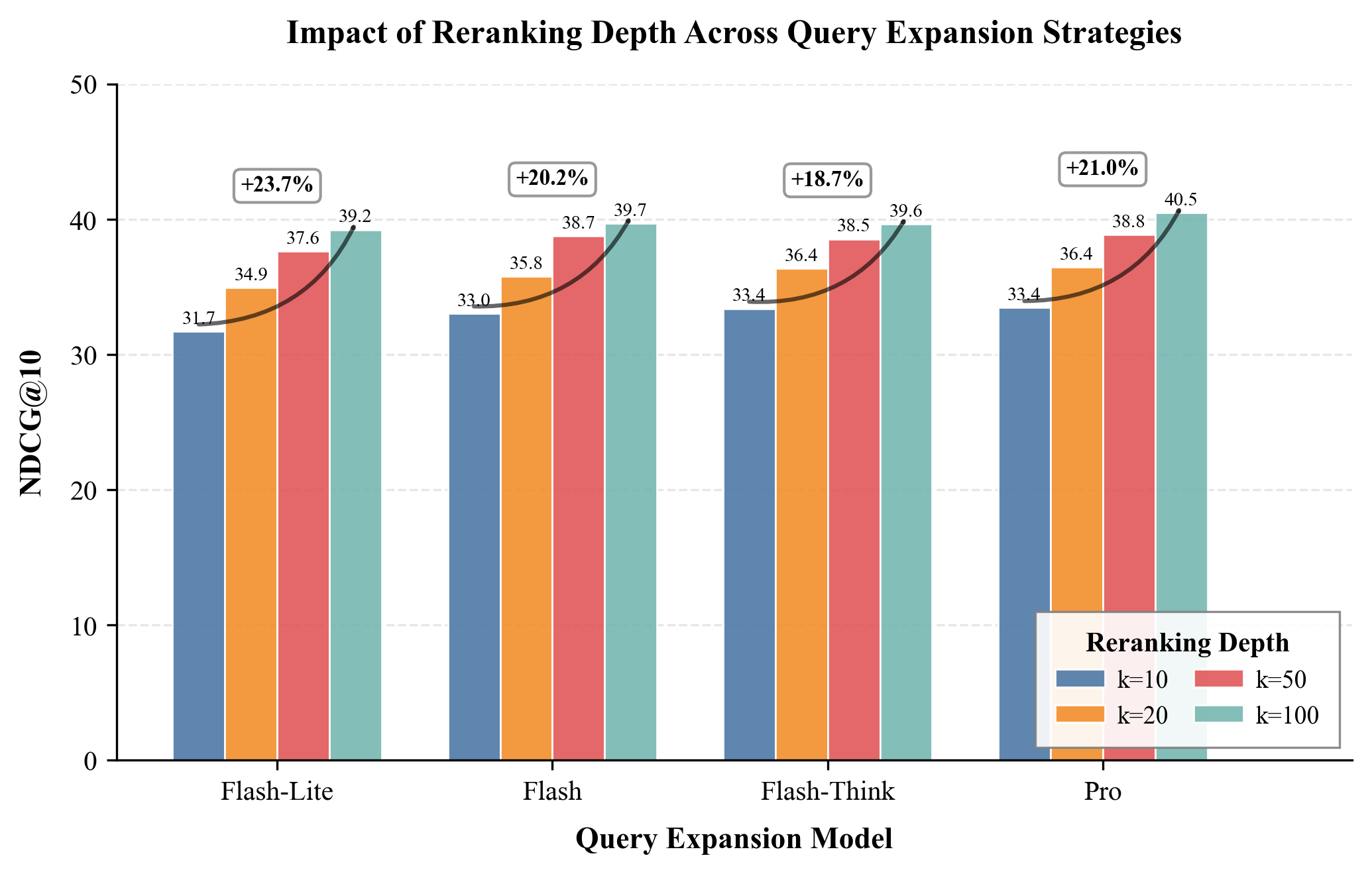}
\caption{Impact of re-ranking depth across query expansion strategies. Increasing the candidate pool from $k=10$ to $k=100$ consistently improves NDCG@10 by approximately 20\% across all query expansion models, demonstrating that episodic retrieval depth is a high-leverage compute allocation mechanism.}
\label{fig:reranking_depth_analysis}
\end{figure}

Figure~\ref{fig:reranking_depth_analysis} Re-ranking depth shows consistent and substantial returns across all QE model variants. Expanding from $k=10$ to $k=100$ yields approximately 20\% relative improvement in NDCG@10 (33 → 40), with gains visible at each increment. Although cost scales linearly with $k$, NDCG@10 improvements remain significant even at the highest values tested. This contrasts sharply with QE, where NDCG@10 plateaus. Allocating compute to deeper validation over a broader candidate set can empirically yield more consistent gains than further scaling the query interface alone.

\section{Conclusion}

RIIR demands substantial LLM compute to bridge the gap between under-specified queries and implicitly relevant documents. This challenge is further amplified as the agent memory stores grow with long-horizon operation. Together, these results support a resource-efficient memory design: use lightweight components for query generation and concentrate compute on deeper, higher-capacity re-ranking, achieving near-maximal quality at substantially lower cost. 

\section{Future Work:} Our study focuses on a single benchmark (BRIGHT) and a single model family (Gemini 2.5) using a single retrieval algorithm. Future work should validate these compute allocation principles across diverse agent memory workloads, alternative model families, and retrieval architectures (e.g., dense retrievers, hybrid systems) to determine whether the asymmetric returns we observe generalize beyond reasoning-intensive scientific queries.

\bibliography{iclr2026_conference}
\bibliographystyle{iclr2026_conference}

\end{document}